\documentclass[aps,pre,onecolumn,groupedaddress,nolongbibliography]{revtex4-2}

\usepackage{graphicx}
\usepackage{xcolor}
\usepackage{amssymb}
\usepackage{ulem}

\definecolor{NewBlue}{RGB}{13,8,135}
\definecolor{NewViolet}{RGB}{132,5,167}
\definecolor{NewRed}{RGB}{211,81,113}
\definecolor{NewOrange}{RGB}{252,166,54}

\begin{document}

\title{Downslope granular flow through a forest of obstacles}
\author{Baptiste Darbois Texier, Yann Bertho, Philippe Gondret}
\affiliation{Universit\'e Paris-Saclay, CNRS, Laboratoire FAST, 91405 Orsay, France.}

\begin{abstract}
We investigate the effect of a forest of pillars on a granular layer steadily flowing over a rough inclined plane. We quantify experimentally how the steady flow rate of grains is affected by the inter-pillars distance for different layer thicknesses and slope angles. We then propose a model based on a depth-average approximation associated with $\mu(I)$ rheology that considers the additional force exerted by the pillars on the granular layer. This model succeeds in accounting for most of the observed results when taking into account some inertia due to the nonvanishing Froude number of the flow.
\end{abstract}

\maketitle

\section{Introduction}
The inclined plane is a classical configuration to investigate the rheological properties of a flowing material. In steady conditions, it gives access to the unknown bottom friction that balances the known gravity force. This configuration has been proved to be suitable for studying the rheology of yield stress fluids \cite{coussot1995determination}, dense suspensions \cite{bonnoit2010inclined} and dry granular materials \cite{GdrMidi2004,pouliquen1999scaling}. For dry granular matter, experiments on an incline show that steady flows only develop between two critical inclination angles $\theta_1$ and $\theta_2$ \cite{daerr1999two,pouliquen1999scaling}. For a slope less than $\theta_1$ there is no flow, and for a slope greater than $\theta_2$ the flow is accelerated. Between these two limits, steady flows are well described by the $\mu(I)$ rheology \cite{jop2006constitutive} which relates the local friction coefficient $\mu$ to the local velocity gradient taking into account the local pressure through the inertial number $I$ \cite{forterre2008flows,andreotti2013granular}. Even if this rheological model suffers from some weaknesses \cite{barker2015well}, this local rheology succeeds in explaining most of the observed results in different geometries and complex flow configurations such as the collapse of a granular column \cite{lacaze2009axisymmetric,lagree2011granular} or the flow around a cylinder \cite{seguin2016clustering}. The knowledge of the rheological behavior of granular matter is crucial for both industrial \cite{gaudel2016granular,gaudel2019effect} and natural \cite{delannay2017granular} situations. In particular, the understanding of granular flows is a first step towards a better prediction of snow avalanches as well as other hazardous geophysical granular flows such as debris flows, lahars and pyroclastic flows. The presence of natural or artificial obstacles is known to reduce the destructive power of these events \cite{feistl2014observations,teich2014computational,yan2021experimental,vedrine2022detrainment,liang2023revealing}. Thus, the physical description of the interaction between granular avalanches and obstacles is a key input for mitigation of various geophysical phenomena. In order to develop a precise understanding of the avalanche-obstacles interaction, several experimental and numerical works have investigated configurations where obstacles are placed in the path of a noncohesive granular flow on an inclined plane. Faug \textit{et al.} have inspected numerically the case of grains flowing down an incline and impacting a wall normal to the incoming flow \cite{faug2009mean}. They showed that the wall generates a triangular stagnant zone upstream that sharply increases the mean force experienced by the wall at low inertial numbers. Cui and Gray investigated the gravity-driven granular flow around a circular cylinder in the supercritical regime \cite{cui2013gravity}. In this situation, they observed that a very sharp bow shock wave is generated in front of the cylinder and a grain-free region forms on the lee side. They studied the position of the bow shock and the granular vacuum as a function of the flow properties and they proposed a simple two-parameters depth-averaged avalanche model to capture their observations.

Other studies have moved beyond the case of a single obstacle and considered the effect of multiple obstacles. Benito \textit{et al.} have investigated the role of a forest of cylindrical pillars on the stability of a granular layer on an incline \cite{benito2012stability}. They revealed that the presence of pillars increases the stability of the granular layer towards larger slope angles. They succeeded in rationalizing this effect by a model that takes into account the additional friction
force exerted by the pillar forest onto the granular layer. More recently, Luong \textit{et al.} have studied the spreading of a mass of grains on a slope through a regular array of pillars \cite{luong2020spread}. They found that the presence of pillars slows down the spreading of the granular mass and enhances its lateral dispersion. Luong \textit{et al.} proposed an empirical model to capture the slow-down of the granular mass using an effective friction coefficient that depends on inter-pillars distance. However, the stationary flow of grains through a forest of obstacles has not been considered yet while it is of interest to better understand the coupling between granular flows and multiple obstacles, and to test the ability of a continuum rheological model to account for the observed results.

This paper presents experimental results and modeling of the stationary granular flow that establishes down an inclined plane where a forest of obstacles is present. Section~\ref{Sec:Experiments} details the setup used to study this steady gravity-driven granular flow and presents the experimental results for the flow rate as a function of pillar density, layer thickness, and slope angle. In Sec.~\ref{Sec:Model}, a modeling based on a depth-average approach associated with a $\mu(I)$ rheology that accounts for the additional resistive force exerted by the pillar forest is shown to succeed in explaining the experimental results. A final discussion is then proposed in Sec.~\ref{Sec:Discussion}.

\section{Experiments}
\label{Sec:Experiments}
\subsection{Setup}
\begin{figure}[t]
\center
\includegraphics[width=\hsize]{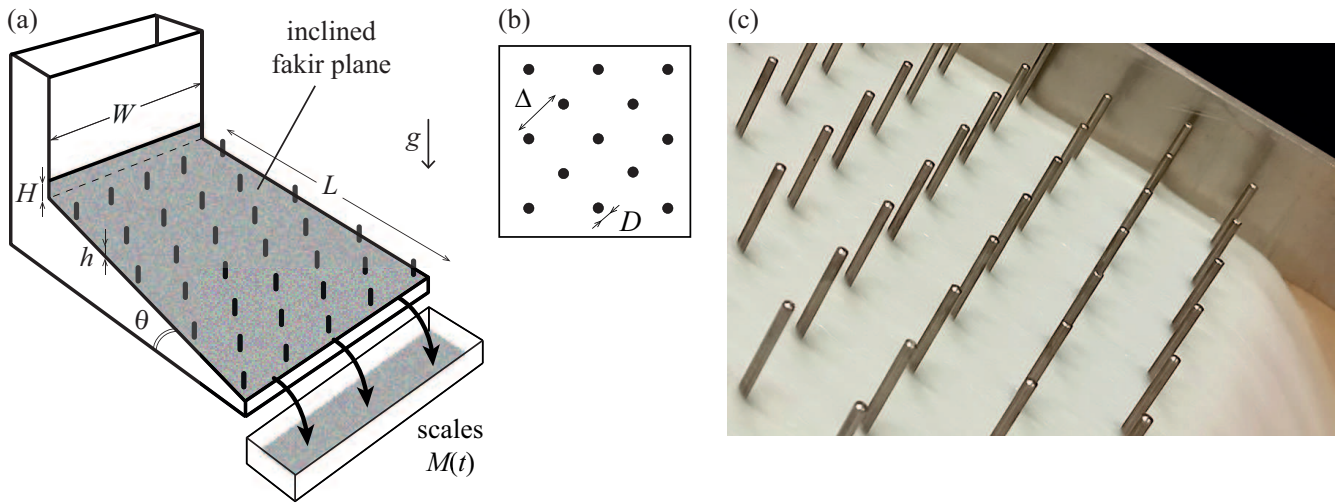}
\caption{(a) Sketch of the experimental setup and notations used. (b) Top view of the inclined plane with the organization of the forest of pillars along a centered square lattice. (c) Picture of the granular medium flowing through a pillar array of separating distance $\Delta = 20$~mm.}
\label{Fig01}
\end{figure}

The experimental setup sketched in Fig.~\ref{Fig01}(a) consists of a plane of length $L=50$~cm and width $W=37$~cm that can be inclined by an angle $\theta$ from the horizontal up to $45^\circ$. The peculiarity of this setup is to include a forest of cylindrical pillars positioned regularly (according to a centered square lattice) perpendicular to the plane surface with a separating distance denoted $\Delta$ [Fig.~\ref{Fig01}(b)]. The diameter of the pillars is $D=2$~mm and their height (35~mm) is always larger than the thickness of the flowing granular layer. Three different forests have been built with $\Delta = 20$, $14$, and $10$~mm corresponding to three pillar densities $\chi=1/\Delta^2$ equal to $0.25$, $0.51$, and $1.00$~cm$^{-2}$, respectively. The null pillar density ($\chi=0$) corresponds to a plane without any pillars. The surface fraction occupied by the pillars, $\phi_s =(\pi/4)(D/\Delta)^2$, is equal to about $0.8$, $1.6$, and $3.1\%$, respectively, for the three different forests, far from the maximal value $\phi_s =\pi/4\simeq 78.5 \%$ corresponding to the limit case $\Delta=D$. The particles used in the experiments are sieved glass beads (density $\rho_g = 2520$~kg~m$^{-3}$) of diameter $d=450$~$\mu$m with a relative dispersion in diameter of order 10\%. The same glass beads have been used to cover the bottom of the inclined plane to ensure a no-slip condition. The cylinder-grain size ratio $D/d$ equals 4.4 and the number of grains between two pillars $(\Delta -D)/d$ is always larger than 17, which prevents clogging \cite{zuriguel2005jamming}.

An experiment consists in filling with grains the reservoir placed at the top of the incline and opening the gate with an aperture $H$ at the bottom of the reservoir. Using the scale placed at the end of the incline, we measure the mass of grains $M$ falling from the plane as a function of time $t$ from which we derive the instantaneous mass flow rate $Q=dM/dt$. A low-incident laser sheet and a camera are used to estimate the mean thickness $h$ of the flowing layer with an accuracy of $0.1$~mm. The thickness of the layer is measured at the center of the slope, away from the proximity of the pillars, and averaged over the longitudinal extent of the laser sheet, which is about 5 cm. It is observed that the thickness of the layer remains constant for the duration of the steady state. We first realize a set of 60 experiments where the slope is constant $\theta = 29.5^\circ$ and where the mean layer thickness is varied between 5 and 22 times the grain diameter. Then, we realize 45 experiments where the inclination is varied between $24.5^\circ$ and $32.5^\circ$ by steps of $0.5^\circ$ and the layer thickness is maintained approximately constant around ten grain diameters. In order to maintain a constant thickness, we manually adjust the gate opening for each inclination angle to compensate for the fact that the average thickness of the layer decreases as the slope increases. For all the experiments, no significant variation of the flowing layer thickness along the transverse and lateral directions of the plane has been observed during the steady regime (within experimental uncertainties) as long as the measurements were not made in the vicinity of the gate or the exit of the inclined plane where entry and exit effects occur. We also measure the velocity of the grains at the surface of the flowing layer and we do not observe any significant variation along the transverse direction, except in a thin layer near the sidewalls ($\approx 3$~mm wide). Neglecting the presence of this layer on the estimate of the flow velocity averaged over the layer thickness, $U=Q/\rho\, W\, h$, leads to a relative error less than 1\%, where $\rho =\phi\, \rho_g = 1510$~kg~m$^{-3}$ is the bulk density of the flowing material and $\phi$ the grain packing fraction. The volume fraction $\phi$ is estimated by weighting the grains contained on a given surface averaged on the thickness of the flowing layer $h$. We find that the averaged volume fraction is constant, $\phi = 0.60 \pm 0.01$, and does not show a clear dependency on the slope inclination nor on the thickness of the flowing layer in agreement with the observations of Benito \textit{et al.}~\cite{benito2012stability}.

Note that our experimental device may be reminiscent of the ``Galton board'' that has been used to study transverse dispersion of grains along their falling path through a forest of obstacles \cite{bruno2003dispersive,benito2007exit}. However, our setup presents several differences with the Galton board: a continuous injection of grains over the whole width of the plane, a large inter-obstacles distance, and a significant friction of the grains with the walls.

\subsection{Results}
\label{sec:result}
\begin{figure}[t]
\center
\includegraphics[width=0.5\hsize]{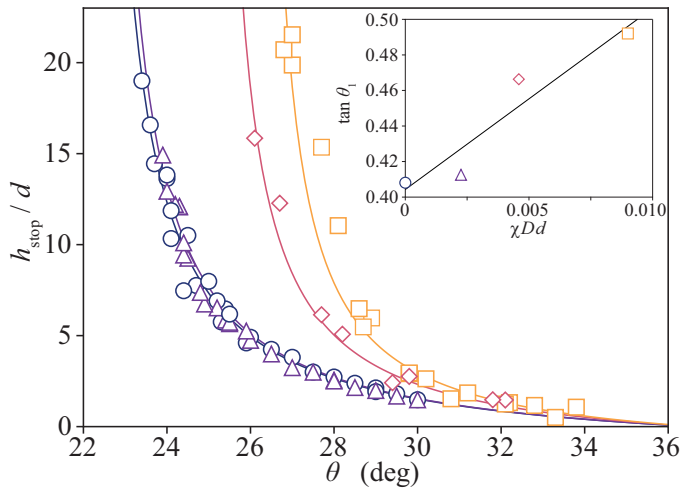}
\caption{Normalized thickness $h_{\rm{stop}}/d$ of the remaining layer after an avalanche as a function of the slope angle $\theta$ for different pillar densities (\textcolor{NewBlue}{$\circ$})~$\chi=0$~cm$^{-2}$, (\textcolor{NewViolet}{$\vartriangle$})~$\chi=0.25$~cm$^{-2}$, (\textcolor{NewRed}{$\diamond$})~$\chi=0.51$~cm$^{-2}$, and (\textcolor{NewOrange}{$\square$})~$\chi=1.00$~cm$^{-2}$. The solid lines are the best fits of the data by Eq.~(\ref{eq:h_stop}) with $\lambda=1.55$, $\theta_2 = 36.6^\circ$, and $\theta_1 = 22.2^\circ,\, 22.4^\circ,\, 25.0^\circ$, and $26.2^\circ$. Inset: Variation of $\tan\theta_1$ with the pillar density $\chi$. The solid line is the best linear fit through the data of equation $\tan\theta_1 = 0.404+10.2\,\chi Dd$.}
\label{Fig02}
\end{figure}

In order to characterize our system, we first measure the thickness $h_{\rm{stop}}$ at which the granular flow stops as a function of $\theta$ for different pillar densities, following the procedure described by Pouliquen~\cite{pouliquen1999scaling}. This critical thickness is the thickness below which no steady flow can be observed. Figure~\ref{Fig02} shows the stopping curves $h_{\rm{stop}} (\theta)$ for the different pillar densities $\chi$ considered in this paper.
We observe that the presence of pillars shifts the curve towards larger slope angles for larger $\chi$, in agreement with the observations of Benito \textit{et al.}~\cite{benito2012stability}. According to \cite{benito2012stability}, these results can be fitted by the equation
\begin{equation}
\frac{h_{\rm{stop}} (\theta, \chi) }{d} =\lambda \, \frac{\tan \theta_2 - \tan \theta}{\tan \theta - \tan \theta_1(\chi)},
\label{eq:h_stop}
\end{equation}
where $\theta_1$ and $\theta_2$ correspond to the characteristic stopping angles of the granular layer of infinite height ($h\rightarrow \infty$) and vanishing layer thickness ($h \rightarrow 0$), respectively, and $\lambda$ is a numerical prefactor characterizing the influence of the rough bottom surface on the stopping height. Note that in Eq.~(\ref{eq:h_stop}), the dependence of $\tan \theta_1$ with $\chi$ results from the additional friction of the granular layer on pillars and does not correspond to a change in the bulk properties of the medium. For the case without pillars ($\chi=0$), the best fit of the data is found for $\theta_1 = 22.2^\circ$, $\theta_2 = 36.6^\circ$, and $\lambda=1.55$. For the cases with pillars ($\chi \neq 0$), we keep $\theta_2$ and $\lambda$ unchanged following the findings of Benito \textit{et al.} who show that only $\theta_1$ depends significantly on $\chi$ \cite{benito2012stability}. The best fits of the curves for the three pillar densities $\chi=0.25$, $0.51$, and $1.00$~cm$^{-2}$ are found for $\theta_1 = 22.4^\circ$, $25.0^\circ$, and $26.2^\circ$, respectively. These fits appear as solid lines in Fig.~\ref{Fig02} and pass quite well through the data. As expected, increasing $\chi$ increases $\theta_1$ as shown in the inset of Fig.~\ref{Fig02}. A linear fit of the form $\tan\theta_1(\chi) = \tan\theta_1(0) + \alpha \chi Dd$, with $\alpha \simeq 10$, passes rather well through the data and is in relatively good agreement with the previous findings of Benito \textit{et al.}~\cite{benito2012stability}.

\begin{figure}[t]
\includegraphics[width=\hsize]{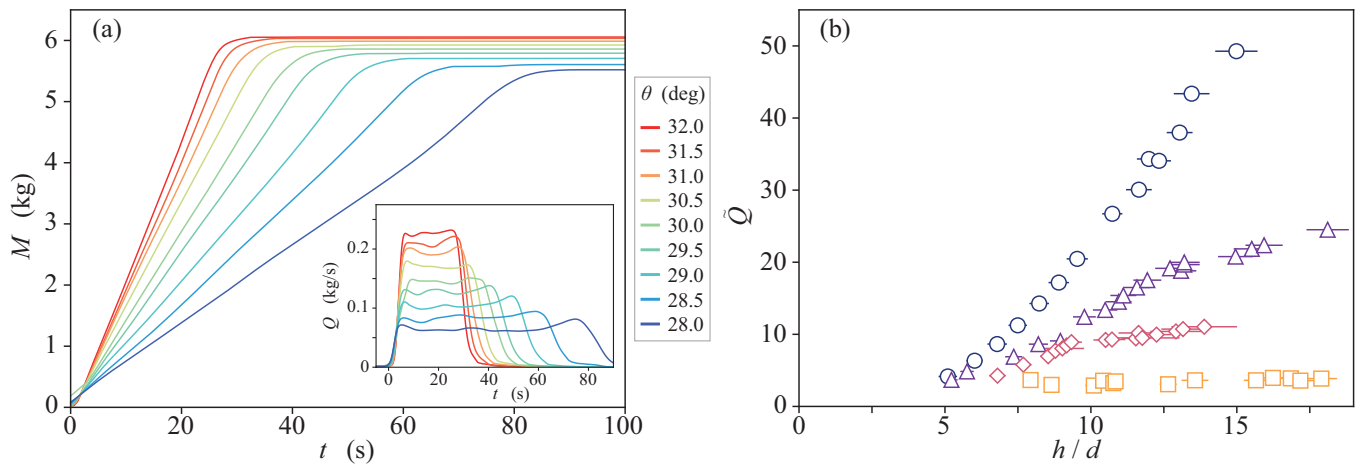}
\caption{(a)~Mass of grains $M$ as a function of time $t$ after opening the reservoir at a given gate aperture ($H=5.5$~mm), a pillar density ($\chi=0.51$~cm$^{-2}$), and nine different slope angles from $28.0^\circ$ to $32.0^\circ$ by steps of $0.5^\circ$. Inset: derived mass flow rate $Q=dM/dt$ as a function of time. (b)~Normalized steady flow rate $\tilde{Q} = \langle Q \rangle/\rho W g^{1/2} d^{3/2}$ as a function of the normalized layer thickness $h/d$ at slope angle $\theta=29.5^\circ$, for different pillar densities (\textcolor{NewBlue}{$\circ$})~$\chi=0$~cm$^{-2}$, (\textcolor{NewViolet}{$\vartriangle$})~$\chi=0.25$~cm$^{-2}$, (\textcolor{NewRed}{$\diamond$})~$\chi=0.51$~cm$^{-2}$, and (\textcolor{NewOrange}{$\square$})~$\chi=1.00$~cm$^{-2}$.}
\label{fig:h_Q}
\end{figure}

Thereafter, we focus on the granular flow that establishes during a discharge of the reservoir. Figure~\ref{fig:h_Q}(a) presents typical evolution of the mass of grains $M$ as a function of time $t$ after opening of the reservoir at a given aperture of the gate ($H=5.5$~mm), a given pillar density ($\chi=0.51$~cm$^{-2}$), and different slope angles ($28^\circ \leq \theta \leq 32^\circ$). The inset of Fig.~\ref{fig:h_Q}(a) shows the instantaneous mass flow rate $Q(t)=dM/dt$ derived from these measurements. After a transient regime, for $t \gtrsim 5$~s, we observe that the mass flow rate quickly increases up to a constant plateau value that is maintained for most of the discharge duration before vanishing when the reservoir is empty. The flow rate from the reservoir is kept approximately constant since the pressure at the outlet does not depend on the height of grains in the reservoir due to Janssen effect \cite{bertho2003dynamical}.
In the following, we focus on the steady flow rate $\langle Q \rangle$ developing in between these two starting and stopping transients and its dependency with the granular layer thickness and the slope angle. Each experiment has been reproduced at least three times in order to determine the average value of the steady flow rate and to estimate the dispersion of the results. Figure~\ref{fig:h_Q}(b) presents the mean normalized flow rate $\tilde{Q} = \langle Q \rangle/\rho W g^{1/2} d^{3/2}$ as a function of the normalized thickness of the flowing layer $h/d$ for a given slope angle $\theta=29.5^\circ$. In the absence of pillars ($\chi=0$), a supra-linear increase of $\tilde{Q}$ with $h/d$ is observed. The presence of the pillar forest ($\chi \neq 0$) reduces the granular flow rate and the trend is no longer supra-linear but rather sub-linear. The reduction of the flow rate is more significant as the thickness of the granular layer or the density of pillars increases. For the smallest pillar density ($\chi=0.25$~cm$^{-2}$), the flow rate is only significantly reduced for large enough layer thicknesses ($h/d \gtrsim 7$), which is consistent with the fact that the stopping layer $h_{\rm{stop}}$ at a slope angle of $\theta=29.5^\circ$ is not significantly increased by a pillar forest of such a low density. Interestingly, the addition of pillars not only reduces the mass flow rate but also changes the dependence of $\tilde{Q}$ with the layer thickness in a nontrivial way. In particular, for $\chi=1.00$~cm$^{-2}$, the flow rate is almost constant and no longer depends on the thickness of the granular layer [Fig.~\ref{fig:h_Q}(b)].

\begin{figure}[t]
\includegraphics[width=\hsize]{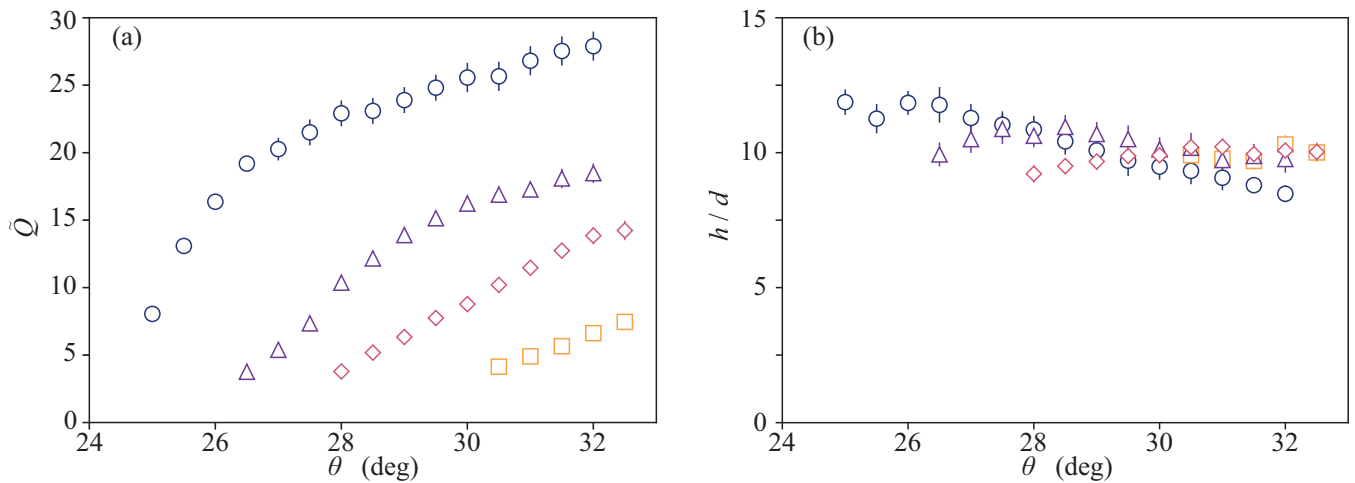}
\caption{(a)~Normalized steady flow rate $\tilde{Q} = \langle Q \rangle/\rho W g^{1/2} d^{3/2}$ and (b)~normalized layer thickness $h/d$, as a function of the slope angle $\theta$ for different pillar densities (\textcolor{NewBlue}{$\circ$})~$\chi=0$~cm$^{-2}$, (\textcolor{NewViolet}{$\vartriangle$})~$\chi=0.25$~cm$^{-2}$, (\textcolor{NewRed}{$\diamond$})~$\chi=0.51$~cm$^{-2}$, and (\textcolor{NewOrange}{$\square$})~$\chi=1.00$~cm$^{-2}$.}
\label{Fig04}
\end{figure}

Then we measure the steady flow rate $\langle Q \rangle$ as a function of the slope angle while keeping the flowing layer thickness constant in the best possible way. Indeed, the layer thickness can hardly be maintained strictly constant because the relation between the gate opening and the granular layer thickness is nontrivial as pointed out by Cui and Gray~\cite{cui2013gravity}. Figures~\ref{Fig04}(a) and \ref{Fig04}(b) show the simultaneous measurements of $\tilde{Q}$ and $h/d$ as a function of the slope angle $\theta$. The normalized flow rate is observed to increase with $\theta$ above a threshold angle which is about $24^\circ$ in the absence of pillars ($\chi=0$). This threshold value corresponds to the value $\theta_{\rm{stop}}(h/d)$ for the forced thickness $h/d \simeq 10$ that is kept approximately constant for different $\theta$ [Fig.~\ref{Fig04}(b)]. This observation is consistent with Fig.~\ref{Fig02} where the flow stops for $h/d \simeq 10$ when $\theta= 24^\circ$ and $\chi=0$. When the pillar density is increased, it induces a shift of the threshold angle towards larger values up to about $30^\circ$ for the highest density ($\chi=1.00$~cm$^{-2}$). Note that this increase of the flow rate is mainly due to an increase in velocity as the thickness of the flowing layer is approximately constant in the whole range of slope angle $\theta$ for each value of $\chi$ [Fig.~\ref{Fig04}(b)].

\begin{figure}[t]
\center
\includegraphics[width=0.5\hsize]{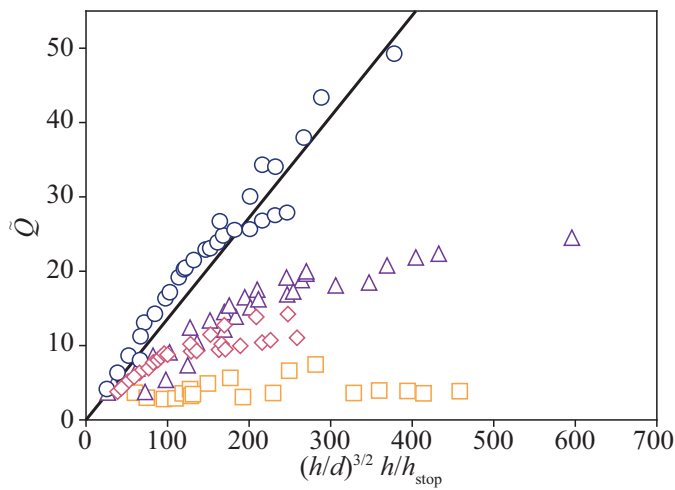}
\caption{Normalized steady flow rate $\tilde{Q} = \langle Q \rangle/\rho W g^{1/2} d^{3/2}$ as a function of the normalized thickness $(h/d)^{3/2}\, h/h_{\rm{stop}}(\chi)$ for different pillar densities (\textcolor{NewBlue}{$\circ$})~$\chi=0$~cm$^{-2}$, (\textcolor{NewViolet}{$\vartriangle$})~$\chi=0.25$~cm$^{-2}$, (\textcolor{NewRed}{$\diamond$})~$\chi=0.51$~cm$^{-2}$, and (\textcolor{NewOrange}{$\square$})~$\chi=1.00$~cm$^{-2}$. The solid line corresponds to the linear trend with the slope of 0.14 proposed by Pouliquen~\cite{pouliquen1999scaling} for glass beads flowing down a slope without pillars.}
\label{Fig05}
\end{figure}

For granular flows down a rough incline without any pillar forest ($\chi = 0$), the mean velocity $U$ normalized by $(gh)^{1/2}$ has been observed to scale linearly with the thickness $h$ normalized by the stopping thickness $h_{\rm{stop}}$ \cite{pouliquen1999scaling,forterre2003long,andreotti2013granular}. This implies that the dimensionless flow rate $\tilde{Q}$ should scale as $h^{5/2}/(h_{\rm{stop}}~d^{3/2})$. By testing this scaling with our data (Fig.~\ref{Fig05}), we notice that the different sets of data obtained for $\chi =0$ follow well such a scaling with the numerical coefficient 0.14 obtained by Pouliquen~\cite{pouliquen1999scaling} for flowing glass beads on a bottom of glued glass beads. However, the data sets corresponding to a pillar forest ($\chi \neq 0$) do not follow such a scaling, or with a slope that decreases significantly as $\chi$ increases. In the next section, we present a model that rationalizes these results.

\section{Model}
\label{Sec:Model}
\begin{figure}[t]
\includegraphics[width=\hsize]{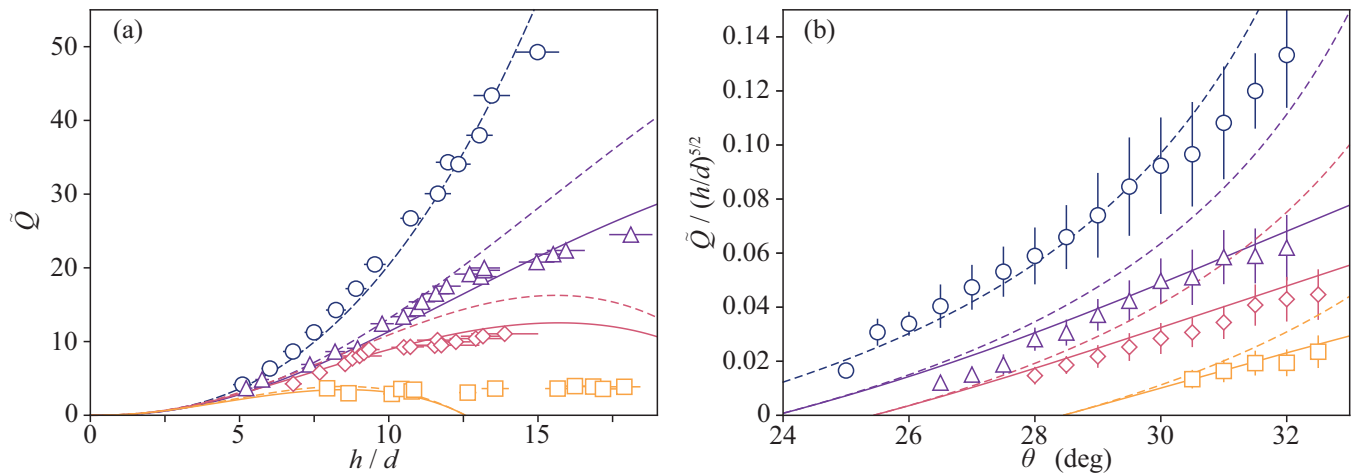}
\caption{(a)~Normalized steady flow rate $\tilde{Q} = \langle Q \rangle/\rho W g^{1/2} d^{3/2}$ as a function of the normalized layer thickness $h/d$ for a given slope angle $\theta=29.5^\circ$. (b)~Normalized steady flow rate $\tilde{Q}$ divided by $(h/d)^{5/2}$ as a function of the slope angle $\theta$. Same data as in Figs.~\ref{fig:h_Q}(b) and \ref{Fig04}. Dashed lines show the noninertial predictions ($C_d=0$) from Eq.~(\ref{eq:model_flow-rate_pillars}) for each pillar density with $I_0=0.35$ and $\beta=1.5$. Solid lines present the inertial numerical solutions of Eq.~(\ref{eq:force_balance_pillars2}) with $\beta=1.5$ and $C_d=3$.}
\label{Fig06}
\end{figure}

Let us now model the granular flow down the incline driven by the gravity $g$ with a depth-average approach by considering the $\mu(I)$ rheology \cite{forterre2008flows}. This rheology relates the friction coefficient of the granular medium $\mu$, defined as the ratio of the shear stress over the normal stress $P$, with the inertial number defined as $I=\dot{\gamma} \, d/\sqrt{P/\rho_g}$, where $\dot{\gamma}$ is the shear rate of the flow. Under the depth-average approximation, a slice of the granular layer experiences its weight and a basal friction dictated by the $\mu(I)$ rheology, two forces that balance in a stationary regime \cite{forterre2003long}. In steady state, the force balance on a slice of the granular layer of thickness $h$ and of unit length and unit width reads
\begin{equation}
\rho g h \sin \theta - \mu_b (I_b) \rho g h \cos \theta =0,
\label{eq:force_balance}
\end{equation}
where the basal friction coefficient $\mu_b$ is a sole function of the inertial number $I_b$ calculated at the base of the granular layer. For a Bagnold velocity profile in the granular layer, the basal shear rate $\dot{\gamma_b}$ is related to the mean flow speed $U=\langle Q \rangle /\rho Wh$ through the relation $\dot{\gamma_b}=5U/2h$. Considering that the pressure at the bottom of the granular layer is $P=\rho gh\cos \theta$, we get that the inertial number at this location is $I_b(h,U)=5\, U\, d/2\, \sqrt{\phi gh^3\cos\theta}$. Experimental results obtained in different flow configurations have permitted the establishment of an expression for the dependency of the basal friction coefficient with flow parameters
\begin{equation}
\mu_b (I_b) = \tan\theta_1 + \frac{\tan\theta_2 -\tan\theta_1}{1 +I_0/I_b},
\label{eq:friction}
\end{equation}
where $I_0$ is the characteristic inertial number above which the friction coefficient increases from $\tan \theta_1$ to $\tan \theta_2$ \cite{andreotti2013granular}. Note that $I_0$ may be viewed as the inverse of a characteristic length scaled by the grain diameter, and depends on the granular material and bed roughness \cite{gray2014depth}. Inserting the friction law Eq.~(\ref{eq:friction}) in Eq.~(\ref{eq:force_balance}), we get a prediction for the normalized flow rate in the absence of pillars:
\begin{equation}
\tilde{Q} (h,\theta) = \frac{2 I_0}{5} \frac{\tan \theta - \tan \theta_1}{\tan \theta_2 - \tan \theta} \sqrt{\phi \cos \theta} \left( \frac{h}{d} \right)^{5/2}.
\label{fig:Q_prediction}
\end{equation}
In order to compare this classical prediction with our observations made in the absence of pillars ($\chi=0$), we present the normalized flow rate $\tilde{Q} = \langle Q \rangle/\rho W g^{1/2} d^{3/2}$ as a function of $h/d$ for $\theta=29.5^\circ$ in Fig.~\ref{Fig06}(a) and the normalized flow rate $\tilde{Q}$ divided by $(h/d)^{5/2}$ as a function of $\theta$ in Fig.~\ref{Fig06}(b), respectively. The blue lines in these plots indicate the best fits of the data with Eq.~(\ref{fig:Q_prediction}) considering the values determined previously for the angles $\theta_1$ and $\theta_2$ (Sec.~\ref{sec:result}) and considering $I_0$ as a free parameter. The best fits are found for $I_0=0.35$ in reasonable agreement with previous estimates made for glass beads ($I_0 \simeq 0.28 - 0.30$) \cite{jop2006constitutive,forterre2008flows}. The model fits quite well the experimental data in Fig.~\ref{Fig06}(a) which means that the dimensionless flow rate $\tilde{Q}$ follows the expected power scaling law $(h/d)^{5/2}$. The model also fits well the data in Fig.~\ref{Fig06}(b) for moderate slope angle $\theta$ but seems to overestimate $\tilde{Q}$ for $\theta \gtrsim 30^\circ$. Note that for a given slope angle in between $\theta_1$ and $\theta_2$, this model predicts a vanishing steady flow for a vanishing thickness $h$ and fails to predict the stopping of the flow at $h_{\rm{stop}}$. This is a known weakness of the $\mu(I)$ rheology as the complex liquid-solid transition of granular matter is not captured by this local approach \cite{andreotti2013granular}.

Subsequently, we develop a model to rationalize the effect of pillars on granular flow. Note that a satisfactory model cannot be achieved by considering different parameters of the $\mu(I)$ rheology. Indeed, such an approach would not capture the change in the dependence of the flow rate on layer thickness and slope angle. For this reason, we consider a new ingredient in the model which is the force applied by the pillars on the granular layer. The force experienced by a single pillar immersed in a granular flow depends on the flow regime which is determined by the Froude number $\mathrm{Fr} = U/\sqrt{gh\cos \theta}$ based on the mean flow speed $U$ and the layer thickness $h$ \cite{faug2015macroscopic}. In the limit of small Froude numbers ($\mathrm{Fr} \ll 1$), the drag force results from the hydrostatic-like pressure, $\rho gz\cos\theta$, applying on the exposed surface of the pillar, $h\,D$, and expresses as $f_h=\beta \rho gh^2 D \cos\theta$, where $\beta$ is a numerical prefactor \cite{albert1999slow}. In the opposite limit of large Froude numbers ($\mathrm{Fr} \gg 1$), the drag force results from inertia and expresses $f_i = C_d\, \rho\, hD\, U^2$, where $C_d$ is a dimensionless drag coefficient.
\begin{figure}[t]
\includegraphics[width=\hsize]{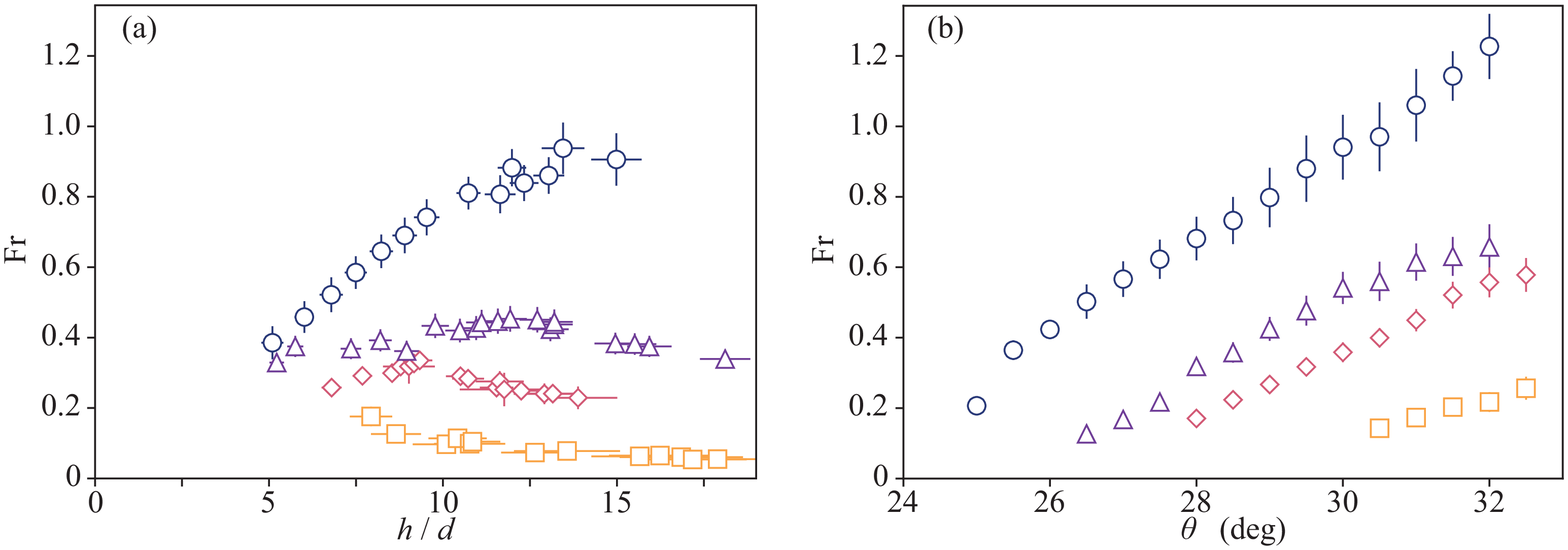}
\caption{Froude number $\mathrm{Fr}$ as a function of (a)~the normalized layer thickness $h/d$ and (b)~the slope angle $\theta$ for four different pillar densities (\textcolor{NewBlue}{$\circ$})~$\chi=0$~cm$^{-2}$, (\textcolor{NewViolet}{$\vartriangle$})~$\chi=0.25$~cm$^{-2}$, (\textcolor{NewRed}{$\diamond$})~$\chi=0.51$~cm$^{-2}$, and (\textcolor{NewOrange}{$\square$})~$\chi=1.00$~cm$^{-2}$.}
\label{Fig07}
\end{figure}
The variations of the Froude number $\mathrm{Fr}$ as a function of $h/d$ and $\theta$ are shown in Figs.~\ref{Fig07}(a) and \ref{Fig07}(b), respectively. Only a few $\mathrm{Fr}$ values are a little larger than 1 and all the $\mathrm{Fr}$ values corresponding to the cases with a pillar forest ($\chi \neq 0$) are smaller than 0.7. In this range, the drag force is expected to correspond mainly to a quasi-static force term with a little inertia. This implies that both contributions to the drag force have to be considered in our model. For simplicity, we will consider that the resulting force in this transitional regime corresponds to the addition of the two limit terms $f_h+f_i$. This crude modeling has been already used to reproduce quite well the variation of the force with $\mathrm{Fr}$ \cite{faug2015macroscopic}, and correctly predicts the dynamics of an object impacting a granular material \cite{seguin2009sphere,katsuragi2013drag}. In order to account for the presence of multiple pillars, we assume that there is no interaction between pillars and that the sum of the individual forces exerted by the pillars on the granular layer results in a global force per unit surface acting on the whole layer of grains which thus simply reads $(f_h+f_i)\chi$. Indeed, the granular flow that establishes in the vicinity of a cylindrical obstacle has been studied experimentally, revealing that the velocity perturbation vanishes on a distance that is roughly one cylinder diameter \cite{seguin2011dense}. In our experiments, the minimal distance between pillars ($\Delta=10$~mm) is much larger than the pillar diameter ($D=2$~mm) so that no interaction between pillars is expected. Under these approximations, the force balance for a slice of the granular layer in the presence of pillars becomes
\begin{equation}
\rho g h \sin \theta - \mu_b (U,h) \rho g h \cos \theta - \beta \rho g \chi h^2 D \cos \theta - C_d \rho \chi h D \, U^2 =0,
\label{eq:force_balance_pillars}
\end{equation}
\noindent which can be written in a dimensionless form as
\begin{equation}
\tan \theta - \mu_b (U,h) - \beta \chi h D - C_d \frac{\chi D \, U^2}{g \cos \theta} =0.
\label{eq:force_balance_pillars2}
\end{equation}
Solving Eq.~(\ref{eq:force_balance_pillars2}) provides the dependency of the mean flow speed $U$ (or equivalently the normalized flow rate $\tilde{Q}$) with the relevant parameters of the problem ($h$, $\theta$, $\chi$, \ldots). Note that in the limit where $\chi \rightarrow 0$, Eq.~(\ref{eq:force_balance_pillars}) leads to Eq.~(\ref{eq:force_balance}) and we recover the prediction of the standard $\mu(I)$ rheology. In the particular case where $C_d=0$, a simple analytical solution of Eq.~(\ref{eq:force_balance_pillars2}) can be found for the normalized flow rate $\tilde{Q}$:
\begin{equation}
\tilde{Q} = \frac{2 I_0}{5} \, \frac{\tan \theta - \tan \theta_1 -\beta \chi hD}{\tan \theta_2 +\beta \chi hD - \tan \theta} \, \sqrt{\phi \cos \theta} \, \left( \frac{h}{d} \right)^{5/2}.
\label{eq:model_flow-rate_pillars}
\end{equation}
In this limit of vanishing inertial drag ($C_d=0$) corresponding to low enough $\mathrm{Fr}$, Eq.~(\ref{eq:model_flow-rate_pillars}) predicts that the presence of pillars changes the discharge flow curve $\tilde{Q}(h,\theta)$ through a shift of both friction coefficients $\mu_1=\tan \theta_1$ and $\mu_2= \tan \theta_2$ (associated to the inclined plane without pillars) by a quantity $\beta \chi hD$ that is directly proportional to the density $\chi$ of pillars and their surface $hD$ exposed to the flow. At a given angle, this correction mitigates the increase of the flow rate with $h$ in qualitative agreement with the observations presented in Fig.~\ref{fig:h_Q}. The shift of $\tan \theta_2$ predicted by Eq.~(\ref{eq:model_flow-rate_pillars}) in the case of a flowing layer contrasts with the fact that no variation of $\tan \theta_2$ with $\chi$ is observed in the case of a stopping layer (as discussed in Sec.~\ref{sec:result}). This difference can be understood from the competition between the friction of the granular layer with the bottom wall and with pillars: indeed, when $\theta$ approaches $\theta_2$, the stopping granular layer necessarily has a negligible thickness and its behavior is dominated by the friction with the bottom wall ($\beta \chi h_{\rm stop}D \ll \mu_b$), whereas for a flowing layer with a larger thickness the resistance of pillars can no longer be neglected ($\beta \chi hD \sim \mu_b$). In order to compare quantitatively the predictions of Eq.~(\ref{eq:model_flow-rate_pillars}) with our data, we used the values determined in the absence of pillars for the parameters $\tan \theta_1$, $\tan \theta_2$, and $I_0$, and we consider the force coefficient $\beta$ as a free parameter. The best fits of our experimental data with the predictions of Eq.~(\ref{eq:model_flow-rate_pillars}) are found for $\beta=1.5 \pm 0.1$, and are shown with dashed lines in Figs.~\ref{Fig06}(a) and \ref{Fig06}(b) for all the pillar densities investigated in this paper. The predictions of the noninertial model capture the main trend in the data, i.e., the lowering of the granular flow with increasing pillar density. Nonetheless, the assumption of null inertial drag ($C_d=0$) is too strong to rationalize finely our experimental observations, in particular the variation of the dimensionless flow rate $\tilde{Q}$ with the slope angle $\theta$ as observed in Fig.~\ref{Fig06}(b).

Let us now consider the general case where the inertial contribution is taken into account ($C_d \neq 0$). For the sake of simplicity, we do not make explicit the analytical solution of Eq.~(\ref{eq:force_balance_pillars2}) but we present its numerical solutions. In order to compare these predictions with our observations, we imposed the values determined in the absence of pillars for the parameters $I_0$, $\tan \theta_1$, and $\tan \theta_2$ and we consider the two force coefficients $\beta$ and $C_d$ as free parameters and adjust them to get the best fit between experiments and theory. The best fits are found for $\beta=1.5 \pm 0.1$ and $C_d=3 \pm 1$, and are presented with solid lines in Figs.~\ref{Fig06}(a) and \ref{Fig06}(b). Accounting for the inertial contribution of the pillar drag permits us to have a much better agreement between theory and experiments, especially for the dependency of the normalized flow rate with the slope angle. Our value of the quasi-static coefficient, $\beta \simeq 1.5$, is rather close to the value $\beta=2.4$ found by Albert \textit{et al.}~\cite{albert1999slow} in the limit where the object size is large compared to grain diameter ($D/d>3$). The difference between these two values remains in the usual range of incertitude for force coefficients associated to quasi-static granular flows \cite{faug2015macroscopic}. Concerning the drag coefficient, our value $C_d \simeq 3$ is in very good agreement with the range $C_d=3 \pm 2$ reported by Faug~\cite{faug2015macroscopic}. The ratio of the inertial and the quasi-static drag forces on pillars can be written as $f_i/f_h = (C_d/\beta)\,\mathrm{Fr}^2$. As $C_d/\beta \simeq 2$, the transition between these two regimes is expected to be around the critical Froude number $\mathrm{Fr}_c = (\beta/C_d)^{1/2} \simeq 0.7$. This confirms that our data obtained with a pillar forest ($\chi \neq 0$) correspond to a quasi-static regime with a little inertia. Nonetheless, the model developed hereby has a major limitation as it predicts that the granular flow of thick layers should be stopped by the largest pillar density for $h/d \gtrsim 12.5$ at $\theta=29.5\,^\circ$ (see the orange solid line in Fig.~\ref{Fig06}(a) corresponding to $\chi=1.00$~cm$^{-2}$), a fact that is not observed in our experiments. This discrepancy highlights the limitation of the spatial averaging assumption. In practice, the pillars have a restricted zone of influence and cannot stop the flow away from their field of action, a behavior that is not accounted for the averaging assumption.

\section{Discussion}
\label{Sec:Discussion}
The model developed in the limit of vanishing inertia ($C_d \simeq 0$ when $\mathrm{Fr} \ll 1$) shows that the presence of pillars is accounted through the parameter $\beta \chi hD$ for a steady flow of thickness $h$. This parameter, which is proportional to $h$, induces a reduction of flow rate that is larger for larger $h$ at a given slope angle $\theta$ and for a pillar forest of given density $\chi$. Note that the stopping thickness $h_{\rm{stop}}$ was found to depend on $\chi$ through the parameter $\alpha \chi D d$ [inset of Fig.~\ref{Fig02}(b)]. Combining the linear variation of $\tan (\theta_1 (\chi))$ with Eq.~(\ref{eq:force_balance_pillars2}) for vanishing $U$ leads to an estimate for the stopping thickness $h_{\rm{stop}}/d =\alpha/\beta \simeq 6$ considering $\alpha \simeq 10$ and $\beta \simeq 1.5$. This estimate is in rather good agreement with our data as we do not observe any steady flow for $h/d \lesssim 5$ [Figs.~\ref{fig:h_Q}(b) and \ref{Fig06}(a)]. According to Eq.~(\ref{eq:model_flow-rate_pillars}), the influence of the array of pillars on the granular flow rate is significant when the parameter $\beta \chi hD$ is of the same order as $\tan \theta - \tan \theta_1$ or $\tan \theta_2- \tan \theta$. In the intermediate case where $\theta=(\theta_1 + \theta_2) /2$, we have $\tan \theta - \tan \theta_1 = \tan \theta_2 - \tan \theta = 0.17$ and the term resulting from the presence of the pillar is equal to one fifth of this value when the pillar density reaches a critical value $\chi_c \simeq 0.02/hD$. Under these conditions, the presence of pillars is expected to reduce the mass flow rate by a factor of two thirds according to Eq.~(\ref{eq:model_flow-rate_pillars}). In the case of the setup studied previously, where the pillar diameter is $D=2$~mm and the layer thickness is about $5$~mm, the critical pillar density above which the presence of pillars impacts significantly the flow rate is $\chi_c = 0.2$~cm$^{-2}$. Transposing the same considerations to an avalanche of granular material occurring through a tree forest with typical values $h\sim 1$~m and $D\sim 0.2$~m, the critical obstacle density should be $0.1$~m$^{-2}$ corresponding to one tree every 3~m.

In this paper, we have explored the effect of a forest of cylindrical pillars on the steady granular flow running down an incline. This experiment constitutes a paradigm for geophysical flow mitigation by a forest of obstacles. In addition to reduce the granular flow rate, the presence of pillars changes drastically the dependency of the flow rate with the thickness of the granular layer, i.e., the discharge flow curve. We accounted for the nontrivial impact of the forest of pillars on the granular flow with a model based on a depth-averaged approach and the $\mu(I)$ rheology. This model includes the effect of pillars through an averaged force applying on the granular layer and neglecting interactions between pillars. This approach has been proved to predict quantitatively the reduction of the granular flow rate induced by the presence of pillars. It can be used to determine the minimal pillar density necessary to reduce significantly the energy carried by the granular flow. These results prove that the $\mu(I)$ rheology can be used to describe nonunidirectional granular flows as the one studied here. Nonetheless, some discrepancies between model predictions and experiments are remaining and should motivate future work to relax the averaging approximation and to study the granular flow occurring in the inter-space between pillars. Indeed,  the velocity field of the grain flow around a cylinder has been already characterized together with the granular temperature field and both have been shown to be localized \cite{seguin2011dense,seguin2013experimental} but it would be interesting to investigate the possible modification of these fields in the case of an incline plane with multiple obstacles. Finally, this paper opens several questions such as the influence of the grain size in relation to the diameter of pillars or the effect of pillar arrangement on the flow. Indeed, if it has already been confirmed that the proposed scaling laws are valid for different grain sizes in the case of a pillar-free plane, it could be interesting to confirm these results in the case of pillar forests. Moreover, it could be useful to consider a random distribution of pillars on the inclined plane, perhaps closer to situations encountered in nature, that should generate local clogging and preferred paths for the granular flow.

\begin{acknowledgments}
We are grateful to P.~Jenffer and G.~Marteau for the conception and realization of the experimental setup, and to J.~Amarni, A.~Aubertin, L.~Auffray, C.~Manquest, and R.~Pidoux for their technical support. This work was supported by ANR PIA Grant No. ANR-20-IDEES-0002.
\end{acknowledgments}

\bibliography{biblio}

\end{document}